\documentclass[journal,comsoc]{IEEEtran}
\usepackage[T1]{fontenc}

\usepackage{tabulary,booktabs}

\usepackage{tabularx}

\usepackage{verbatim} 

\ifCLASSINFOpdf
\else
\fi

\usepackage{graphicx, url}

\usepackage{amsmath}
\usepackage[cmintegrals]{newtxmath}
\begin{document}
%
\title{Range Variation Monitoring \\with Wireless Two-Way Interferometry (Wi-Wi)}
%
%
%

\author{Bhola R. Panta, Kohta~Kido, Satoshi~Yasuda, Yuko Hanado, \\Seiji Kawamura, Hiroshi Hanado, Kenichi Takizawa, Masugi Inoue, and Nobuyasu Shiga 
\thanks{Bhola Panta is with Earthquake Research Institute of the University of Tokyo, 1-1-1 Yayoi, Bunkyo-ku, Tokyo 113-0032, Japan. The other authors are with National Institute of Information and Communications Technology (NICT), 4-2-1 Nukui-Kitamachi, Koganei, Tokyo 184-8795, Japan.
Nobuyasu Shiga is also with PRESTO, Japan Science and Technology (JST), 
4-1-8 Honcho, Kawaguchi, Saitama 332-0012, Japan. Corresponding author email: shiga@nict.go.jp}
}

\maketitle

\begin{abstract}

We demonstrated a simple technique for monitoring range variation with millimeter-precision between two remote sites using off-the-shelf
wireless communication modules. The need for the flexible positioning of wireless devices is significantly increasing as more devices are being connected and new services are being developed that require devices to collaborate with one another. We showed that one can monitor the distance variation by analyzing the propagation delay of the wireless communication signal between devices. We previously reported a technique for synchronizing clocks with picosecond precision by monitoring the time variation of two rubidium clocks located at remote sites. Precise measurement of the propagation time variation was necessary for precise synchronization of the clocks, and we used this information to estimate the distance with high precision. In a localized situation, our technique makes it easy to implement a millimeter-precision measurement system. Furthermore, it is less complex in terms of system design and can be a low-cost alternative to existing systems that require precise position measurement. We envision that this demonstrated protocol will be implemented in wireless communication chips and microprocessing units.
\end{abstract}


%
\IEEEpeerreviewmaketitle



\section{Introduction}
%
%
%
%
\IEEEPARstart{F}{or} precise clock synchronization or ranging, the Global Navigation Satellite System, particularly the Global Positioning System (GPS), has become the de facto standard. However, the history of ranging using radio waves goes back to the British Gee and Decca, 
and American LORAN navigation systems, which used hyperbolic measurement techniques \cite{Scholes1952}, long before GPS became a household name. 

As the need for short-range distance estimation for both outdoors and indoors grew, ranging systems based on the received signal strength (RSS), the time or time difference of arrival (ToA/TDoA), the angle of arrival (AoA), the phase of arrival (PoA) metric were extensively researched and made available. Some of them were proposed or surveyed in \cite{Pahlavan2002}, \cite{Miesen2012}, and \cite{Liu2007}. Reference \cite{Ma2014} compared PoA, and ToA-based ranging behavior for Radio-frequency identification (RFID) application from the viewpoint of the Cramer-Rao lower bound. By using ray tracing as well as taking measurements in a real environment, they reported that for RFID, PoA-based ranging showed better performance than ToA. Indeed, in \cite{Exel2013} a method based on the carrier phase of IEEE 802.11 wireless LAN was proposed and it was reported that the method was capable of sub-centimeter precision. A recent paper \cite{Kotaru2017} also used phase information of the channel state information between Wi-Fi packets to track the position of a virtual-reality headset with sub-millimeter precision.

Precision ranging and positioning systems based on UWB have also been proposed, exhibiting sub-centimeter \cite{Petroff2015} as well as millimeter-level \cite{Zhang2010} precision. More recently, geofencing systems using RSS-based distance estimation with Bluetooth Low Energy have become available, such as Apple's iBeacon and Google's Eddystone, but they are not geared towards precision ranging. The proposed schemes with higher precision also tend to be more complex for practical implementation. For example, in the case of \cite{Exel2013}, the carrier synchronization quality and phase estimation jitter affect the results. For the majority of UWB-based transceivers, despite their simple hardware, most manufacturers are unable to produce inexpensive transceivers \cite{Alarifi2016}. 

Laser range finders (LRFs) are extensively used for precise distance measurements in construction, manufacturing, land surveying, forestry, and so forth. Interferometric LRFs provide better accuracy by measuring the interference between incident and reflected beams. In \cite{Shiga2016}, we introduced the wireless two-way interferometry (Wi-Wi)  technique, which can be considered as an interferometric range variation measurement system using wireless signal.   

The ubiquitousness of GPS for distance measuring applications has already been mentioned. However, several factors must be considered before implementing GPS-based applications. One of the main concerns is the availability of the GPS signal. This means that GPS-based applications can only be used outdoors. Even outdoors, the required number of GPS satellites may not be visible in dense, canyon-like areas at a given time, degrading the performance of applications or making them unusable.   

In the experiment described in this paper, we measured only the phase of the carrier wave, so unlike GPS and LRFs, our experiment does not provide absolute range measurement but the variation of the range from an initial position. Wi-Wi can be used indoors as well as outdoors, including in urban, canyon-like environments. With Wi-Wi, it is possible to implement a precision range-monitoring system at any time, indoors or outdoors, using off-the-shelf components. Considering its lower complexity and cost and ease of use, a system using Wi-Wi has significant benefits.


\section{System Model}
\subsection{Calculating Propagation Time Using Wi-Wi}

In this section, we explain the method of calculating the propagation time, which is based on the two-way time transfer (TWTT)  \cite{Hanson1989} and two-way carrier phase (TWCP) \cite{Fujieda2012} techniques.
\begin{figure}[htb]
\begin{center}
\includegraphics[width=8.0cm]{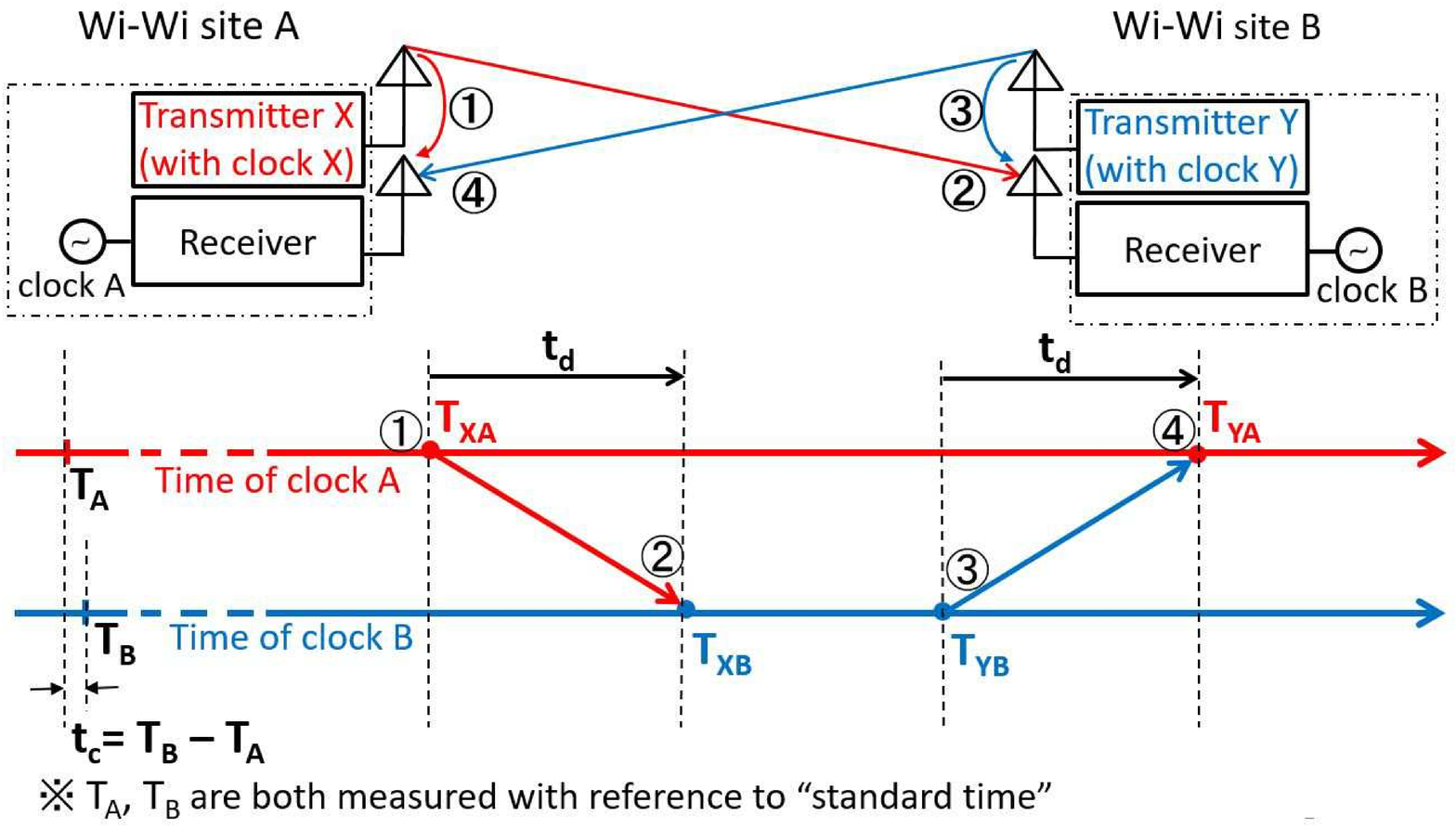}
\end{center}
\caption{System model of Wi-Wi. The upper half shows the setup of the components and  the lower half shows the time sequence.}  \label{fig:system}
\end{figure}

In Fig.~1, the two Wi-Wi systems, site A on the left and site B on the right, have two identical setups. Transmitters X and Y transmit 2.4 GHz Zigbee radio signals, and the signal from each transmitter is received at two sites: at the local site and at the remote site. We exchange the measured phase values between the two sites by including phase information in the subsequent packets. For a detailed description of the time sequence of this procedure, please refer to section 2.5 of  \cite{Shiga2016}.

In the following derivation, we denote time as \textit T and the time interval as \textit t. Let $t_c$ be the time difference between the time of clock B,  $T_B$, and that of clock A, $T_A$, 
i.e.,
\begin{equation}
t_c  \equiv T_B - T_A.
\end{equation}

Let $t_d$ be the propagation time between the two sites.
In this study, transmitter X at site A sends a signal to site B at $T^{}_{XA}$.  After a propagation time $t_d$, the receiver at site B receives the signal at $T^{}_{XB}$. The time difference between the reception at site B and the transmission at site A, $t_B$,  can be expressed as
\begin{equation} \label{eq:tB}
t_B  \equiv T^{}_{XB} - T^{}_{XA} =  t_c +  t_d.
\end{equation}

The effect of the propagation time can be measured using the TWTT technique. Hence, similar to the above procedure, we send the transmission time of site B to site A. Transmitter Y at site B sends the signal to site A at $T^{}_{YB}$. After a propagation time $t_d$, the receiver at site A receives the signal at $T^{}_{YA}$. The time difference between the reception at site A and the transmission at site B,   $t_A$, can be expressed as
\begin{equation}\label{eq:tA}
t_A  \equiv T^{}_{YA} - T^{}_{YB}  =  - t_c +  t_d.
\end{equation}
Note that  the transmission times $T^{}_{XA}$ and  $T^{}_{YB}$ are the reception times measured at sites A and B with clocks A and B, respectively. We assume that the variations in $t_c$ and $t_d$ during the round-trip communication are negligible. 

Adding (2) and (3), we obtain

\begin{equation}\label{eq:td}
t_d  = \frac{t_A + t_B}{2}.
\end{equation}


\subsection{Calculating Distance Using Carrier Phase}
Because Wi-Wi uses the TWCP in this experiment, we rewrite (4) using the phase and derive the distance $l_d$ between the two sites as  

\begin{equation}
l_d \equiv c \cdot t_d = -\left\{{\frac{\phi_A + \phi_B} { 2 \pi}  + K }\right\} \frac{\lambda} { 2},
\end{equation}
where $c$ is the speed of light, $\lambda$ is the wavelength of the carrier wave, $\phi_A$ is the phase advance of the transmitted signal after its propagation from B to A, and $\phi_B$ is the phase advance of the transmitted signal after its propagation from A to B. We note that the negative sign on the right hand is due to the definition of $\phi_A$ and $\phi_B$, according to their measurements at A and B, respectively. Transmitted signal at $t=0$ is received and measured in reference to the receiver's phase at $t=t_d$. As the distance $l_d$ is increased, the measured phase of transmitted signal is shifted negatively. The measured phase can only be obtained modulo 2$\pi$ and $l_d$ can only be obtained modulo $\lambda$/2. The integer ambiguity \textit{K} must be determined separately to obtain the absolute value of $l_d$. However, by measuring $\phi_A$ and $\phi_B$ repeatedly, the variation from the initial measurement can be tracked using the unwrapped phase when the variation of $\phi_A$ and $\phi_B$ between consecutive measurements is guaranteed to be smaller than $\pi$. This principle has been demonstrated in \cite{Fujieda2012}.

The Wi-Wi system (Fig.~1) is configured with four clocks. The goal here is to compare clock A and clock B by eliminating the phase difference  between the transmitter's built-in clock (clock X or clock Y) and the receiver clock at the same site (clock A or clock B). Below, we describe the procedure.  

The phase of clock B with respect to clock A,  $\phi_B$, is measured using transmitter X, and the phase of clock A with respect to clock B,  $\phi_A$, is measured using transmitter Y. They are expressed as 
\begin{equation}
\phi_B\equiv \phi^{}_{XB} - \phi^{}_{XA},
\end{equation}
\begin{equation}
\phi_A\equiv \phi^{}_{YA} - \phi^{}_{YB},
\end{equation}
where $\phi^{}_{XB}$ denotes the phase of clock X received at site B and $\phi^{}_{XA}$ denotes the phase of clock X received at site A. Similarly, $\phi^{}_{YA}$ denotes the phase of clock Y received at site A and $\phi^{}_{YB}$ denotes the phase of clock Y received at site B. A and B denote the locations of reception. Although the phases of clocks X and Y have no correlation with the rubidium clocks owing to their inaccuracies, $\phi_B$ and $\phi_A$ reflect the phase difference between clocks A and B, including the initial offset.

\begin{figure}[htb]
\begin{center}
\includegraphics[width=8.0cm]{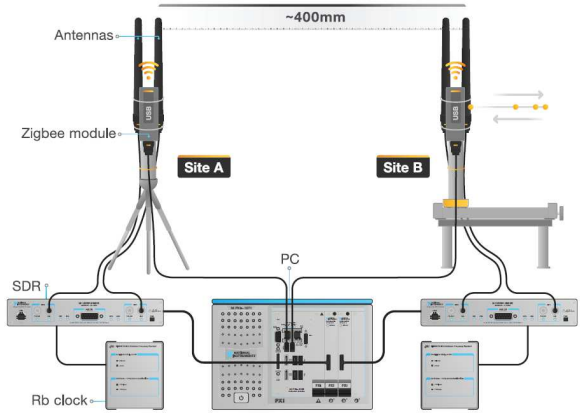}
\end{center}
\caption{Schematic diagram of the experimental setup}  \label{fig:setup}
\end{figure}

\section{Experimental Setup}
Fig. 2 shows a schematic diagram of the set up for an experiment to demonstrate range variation monitoring using the system described in section II. The experiment was conducted in an office with several pieces of furniture and server cabinets. The following paragraph describes the configuration setting, including technical specifications. 

The two sites A and B were separated by about 40 cm. They had an identical setup and each site consisted of a wireless communication module (Zigbee), a software-defined radio (SDR) with two receivers, and a rubidium clock. The specifications of each equipment are summarized  in Table 1.

\begin{table}[h!]
  \centering
  \caption{Specifications of Items Used in the Experiment}
  \label{tab:table1}
\begin{tabularx}{\linewidth}{lX}
    \toprule
    Item name & Specification/Model/Manufacturer \\
    \midrule
    \addlinespace[1ex] Zigbee module & Freq.: 2.4 GHz, Power output: 2.5 dBm \newline Model:TWE-Lite-USB, \newline TOCOS (MONOWIRELESS) \\
    \addlinespace[1ex] Rubidium clock & SIM940, Stanford Reserach Systems\\
    \addlinespace[1ex] Software radio & USRP-2942, National Instruments\\
    \addlinespace[1ex] Personal computer & PXIe-1071, National Instruments \\
    \bottomrule
\end{tabularx}
\end{table}

The Zigbee modules were connected to a PC through USB cables. The SDRs were referenced to 10 MHz, generated by the rubidium clocks, through coax cables.  The PC and SDRs were connected by PCI Express (PCIe) cables. Separate channels were assigned to each Zigbee module at the two locations to avoid channel confusion. The SDR had two receive channels: channel 1 received the signal from the transmitter at the local site and channel 2 received the signal from the remote site. Zigbee communication is based on the IEEE 802.15.4 standard, which employs offset quadrature phase-shift keying (O-QPSK). 

Transmitters X and Y transmitted the signals alternately at a rate of approximately 10 times per second. The SDR digitized the baseband in-phase (I) and quadrature (Q) 16-bit signals at a rate of 4 megasamples/s and the signals were transferred to the PC via the PCIe cable. The PC decoded the signals, symbol by symbol by taking the correlation of the 16 possible chip value sequences. The phase was measured for each symbol and the average of all the unwrapped symbol phases was recorded, which was used to represent the phase of each packet. To demonstrate range
variation monitoring, we changed the position of the antennas of site B by moving the bundled antennas attached to a stepper motor, whose step size was 1~$\mu$m.

A conceptual diagram of the antenna distance variation is shown in Fig.~3. Measurements were taken at positions of 5 mm (b), 7 mm (c), and 8 mm (d) from the initial position (a). Then the measurement direction was reversed as illustrated in the figure. 
\begin{figure}[htb]
\begin{center}
\includegraphics[width=9.0cm]{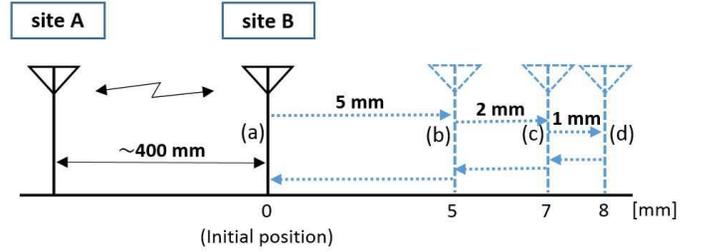}
\end{center}
\caption{Conceptual diagram of the antenna distance variation. (b), (c), and (d) represent distances of 5 mm, an additional 2 mm, and an additional 1 mm from the previous position, respectively. The distance was varied using a stepper motor.}  \label{fig:AntDistVar}
\end{figure}

\section{Results}

Fig.~4 shows the measurement results. We measured the antenna distance variation over a 6 minute interval. The clock difference, plotted in red, varied by about 1 ns over that interval, which is reasonable for rubidium clocks. The timing and distance of the antenna movements are indicated in the magnified plot of $t_d$, in the lower figure. 

In our previous paper \cite{Shiga2016}, we reported a deviation of 2.2~ps in propagation time measurement when the antennas were stationary. This means that we can monitor the variation of the distance between the antennas with 1 mm precision since the wireless signal travels about 1 mm in 0.3~ps. When we changed the distance by 5, 2, and 1 mm, the propagation time changed by 17, 7, and 3~ps, respectively.
\begin{figure}[htb]
\begin{center}
\includegraphics [scale = 0.55] {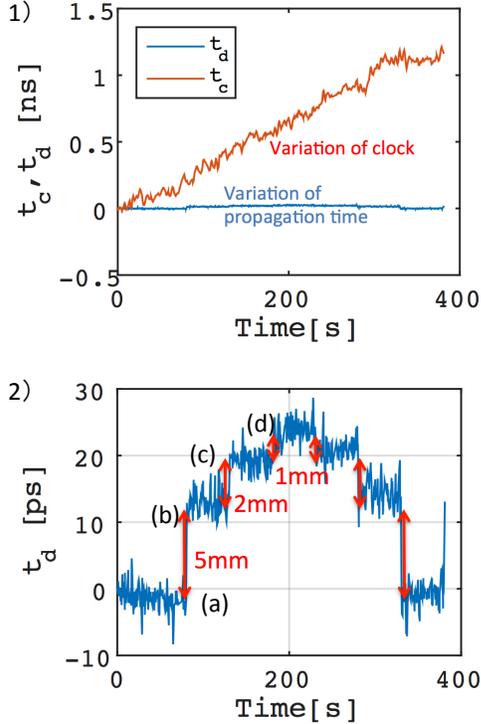}
\end{center}
\caption{Experimental results of Wi-Wi. 1) Variation of clock, $t_c$, and variation of propagation time,  $t_d$, over 6 minute interval. 2) Magnified plot of  $t_d$.  The distance between the antennas was changed using a stepper motor as indicated.}  \label{fig:plot}
\end{figure}

We took the average of the symbol phases over the whole packet to represent the packet phase. Between packets, we used unwrapping  to calculate $t_c$ and $t_d$, as mentioned in the previous section. The unwrapping worked well because the distance variation between adjacent measurements was much smaller than 6 cm, which corresponds to a phase of $\pi$ radian. This method can measure the distance variation only when the variation  between adjacent measurements is guaranteed to be smaller than one half of the wavelength. Note that $t_c$ must be continuous and therefore the stability of the reference clock needs to be much better than 0.5~ns (corresponding to $\pi$ radian) during measurement interval (0.1 s).

\section{Discussion and Conclusion}
In this paper, we demonstrated that the proposed Wi-Wi system can monitor the variation of the distance between two antennas with millimeter precision. The core idea of measuring the distance variation between two points is to monitor the phase variation of transmitted signals in both directions. As long as one can monitor the phase of the transmitted carrier wave, Wi-Wi can achieve range variation measurement. We can apply the method described in this paper to any commercial communication device. So far, we have carried out tests using 2.4 GHz Zigbee, Bluetooth Low Energy, and 920 MHz modules. 

Owing to the development of integrated circuits, many wireless communication chips digitize the I and Q signals of the carrier wave and process the signal digitally to decode the payload. We envision that phase measurement and Wi-Wi measurement can be embedded in wireless chips and microprocessor units, providing opportunities for the extremely low cost and easy implementation of synchronization and distance measurement using any wireless communication scheme.

The Wi-Wi system is based on interferometric phase measurements of the carrier wave. In this regard, our technique closely resembles the principle of interferometric LRFs. However, the LRFs require very precise beam alignment and in practical situations, if the reflector of an LRF is moved, setting up and retuning can be cumbersome. Also, if the reflector is on an inclined surface, the measurement may not be accurate. On the other hand, because we used radio waves, and since radio wave propagation is broad by nature, there is no need for precision beam alignment, thereby making the Wi-Wi system easy to use. In a localized environment such as a construction site, surveying for distance measurement to place markers is critical but often carried out in almost chaotic environment. Therefore, the ability to set up and use a measuring system on-the-fly is highly desirable. 

One of the promising applications of propagation measurement using Wi-Wi is monitoring the distance variation between two points. For example, one can use this technique to monitor small changes  ({\raise.17ex\hbox{$\scriptstyle\mathtt{\sim}$}} mm) in the length or tilt of infrastructure such as towers and bridges. Another interesting application could be atmospheric remote sensing by monitoring the integrated index of refraction (IIR) of air. Using Wi-Wi, the variation of the propagation delay, which is mainly due to water vapor, can be estimated with picosecond precision. 

The proposed method is conceptually simple and is attractive because of its low complexity and straightforward setup to carry out experiments. The current system requires rigorous evaluation in  multipath environments before its practical use. We believe that developing appropriate measures for multipath environment will lead to its wider adoption in academic research as well as in industry.

\section{Future Development}
The prototype version we tested used software radio (USRP) and required AC power, and the overall setup is not conductive to mobile use. Miniaturization of the system is already underway and we have fabricated a module with all the required components (high-precision oscillator, TX/Rx and I/O interfaces), a prototype of which is currently being tested.

\section*{Acknowledgments}
This work was supported by the JST PRESTO program (Grant No. JPMJPR14D5) and by NICT. We thank Ryuichi Ichikawa of NICT for his insightful comments. 

\ifCLASSOPTIONcaptionsoff
  \newpage
\fi


\end{document}